\documentclass[a4paper]{jpconf}

\usepackage{graphicx}

\usepackage{hyperref}

\usepackage{cite}

\newcommand{\maestro}{{\sffamily Maestro}}
\newcommand{\castro}{{\sffamily Castro}}
\newcommand{\starkiller}{{\sffamily StarKiller}}
\newcommand{\starlord}{{\sffamily StarLord}}
\newcommand{\nyx}{{\sffamily Nyx}}
\newcommand{\amrex}{{\sffamily AMReX}}

\newcommand{\cpp}{C\nolinebreak\hspace{-.05em}\raisebox{.4ex}{\tiny\bf +}\nolinebreak\hspace{-.10em}\raisebox{.4ex}{\tiny\bf +}}

\usepackage{color}
\begin{document}

\title{Meeting the Challenges of Modeling Astrophysical Thermonuclear Explosions:
\castro, \maestro, and the \amrex\ Astrophysics Suite}

\author{M. Zingale$^1$,
        A.~S. Almgren$^2$,
        M.~G. Barrios Sazo$^1$,
        V.~E. Beckner$^2$,
        J.~B. Bell$^2$,
        B. Friesen$^{3,2}$,
        A.~M. Jacobs$^4$,
        M.~P. Katz$^5$,
        C.~M. Malone$^6$,
        A.~J. Nonaka$^2$,
        D.~E. Willcox$^1$, and
        W. Zhang$^2$}

\address{$^1$Department of Physics and Astronomy, Stony Brook
  University, Stony Brook, NY 11794-3800 USA}

\address{$^2$Center for Computational Sciences and Engineering,
  Lawrence Berkeley National Lab, Berkeley, CA 94720 USA}

\address{$^3$National Energy Research Scientific Computing Center,
  Lawrence Berkeley National Lab, Berkeley, CA 94720 USA}

\address{$^4$Department of Physics and Astronomy, Michigan State
  University, East Lansing, Michigan 48824 USA}

\address{$^5$NVIDIA Corporation, 2788 San Tomas Expressway,
  Santa Clara, CA, 95050 USA}

\address{$^6$Los Alamos National Laboratory, Los Alamos, NM, 87545 USA}

\begin{abstract}
We describe the \amrex\ suite of astrophysics codes and their
application to modeling problems in stellar astrophysics.
\maestro\ is tuned to efficiently model subsonic convective flows
while \castro\ models the highly compressible flows associated with
stellar explosions.  Both are built on the block-structured adaptive
mesh refinement library \amrex.  Together, these codes enable a
thorough investigation of stellar phenomena, including Type Ia
supernovae and X-ray bursts.  We describe these science applications
and the approach we are taking to make these codes performant on
current and future many-core and GPU-based architectures.
\end{abstract}

\section{Introduction}

Astrophysical explosions come in many flavors: gravitational and
thermonuclear supernovae, unstable burning on the surface of compact
objects, and explosive ignition of burning stages in stellar
evolution.  Accurate modeling of these events requires the coupling of
hydrodynamics, gravity, thermonuclear reactions, and in some cases,
radiation and magnetic fields.  Further, these environments are
characterized by a wide range of length scales, from the size of the
star or binary system down to the burning zone width and dissipation
scales.  Temporal scales are equally impressive---stellar evolution
occurs over 10s of millions to billions of years, the simmering phase
leading up to explosions lasts hours or days to millenia, and the
explosion can be over in seconds to hours.  The radiation leakage,
which leads to the observables we see lasts from hours to months.

No single algorithm meets all of the demands imposed by these events.
Instead, we advance our understanding of these events by piecing
together simulations of different phases of the evolution from
different codes.  Here we discuss our simulation codes, \maestro\ and
\castro, designed to perform three-dimensional models of the early
subsonic evolution leading to runaway and the subsequent explosion,
respectively.  Together this suite of codes allows us to address many
problems in stellar and nuclear astrophysics.  We describe some of the
design details, the current architecture of the code, and some
applications below.

\section{Science drivers and challenges}

Our interests are thermonuclear explosions, including Type Ia
supernovae (SNe Ia), X-ray bursts (XRBs), and novae.  The basic
ingredients for these events are thermonuclear energy release and a
degenerate equation of state that allows a runaway to build without a
pressure response.  Most of the current models for these events are
characterized by a long timescale ``simmering'' phase where reactions
heat the star or layer and drive convection.  Eventually, reactions
become vigorous enough that a runaway takes place, perhaps with an
accompanying burning front that spreads through the star.

\subsection{Type Ia supernovae}

Among the most significant open questions for SNe Ia is the identity of the
progenitor.  About 20 years ago, the community had
mostly converged upon the single-degenerate scenario---a
Chandrasekhar-mass C/O white dwarf that accretes from its companion,
eventually leading to runaway at the center that burns through the
star (see~\cite{hillebrandtniemeyer2000} for the state of the field at
that time).  Since then, a wealth of observations has indicated that
there is a lot of diversity in SNe Ia, and searches for progenitor
systems have strongly suggested that Chandra-mass white dwarfs cannot
explain most SNe Ia.  Today, merging white dwarfs (the double
degenerate scenario) has perhaps become the most popular model.
Other progenitors, like He burning on the surface of a sub-Chandra
white dwarf, have also seen interest in explaining some of the observed
diversity.  See~\cite{araa-maoz} for a review.

There are open questions in all of these scenarios that can be
addressed through simulation.  For the Chandra and sub-Chandra models,
what is the distribution (spatial and temporal) of the hotspots set up
by turbulent convection that give rise to burning fronts?  A
longstanding question with the Chandra model is whether a deflagration
can transition to a detonation during the burning front propagation
through the star.  For the sub-Chandra double detonation model, it is
not clear whether it is possible to create a detonation in the thin
surface He layer.  For double degenerates, it is still unresolved
whether the burning takes place promptly or after some delay.  However
it proceeds, we need to avoid an accretion-induced collapse to a
neutron star.  For many of these investigations, we need to address
numerical issues such as whether it is possible to accurately model
the ignition of a detonation with the spatial resolution we can
attain.  These are some of the questions we seek to answer.

\subsection{X-ray Bursts}

X-ray bursts---the burning of accreted H/He on the surface of a
neutron star---can be important probes of neutron star structure.
Interpreting observations requires that we understand what we are
seeing, which can be influenced by the products of the burning and how
the burning spreads across the star.  Many efforts have focused on
different aspects of these events.  One-dimensional models capture the
energetics well and inform us about the nucleosynthesis
\cite{woosley-xrb}.  Global models show the importance of rotation in
confining the burning \cite{SPIT_ETAL02}, while models inspired from
atmospheric science can explore the vertical structure
\cite{cavecchi:2012}.  However, the resolution differences from the
scale of the burning to a reasonable fraction of the neutron star
surface has prevented detailed explorations of the burning and how it
feeds back on the flame structure and propagation in resolved
calculations.  Algorithms and computing architectures are starting to
reach the point where we can span the gaps in spatial scales between
these calculations to provide a better understanding of the ignition
and propagation of the burning front, and the nucleosynthesis
produced.

\subsection{Requirements}

These problems share common algorithmic requirements: strong coupling
between hydrodynamics and burning, support for a general equation of state,
self-gravity, including isolated boundary conditions, and long timescale
evolution for the convective phases.  All of these problems are inherently
three-dimensional, as turbulence, fluid instabilities, and rotation affect
the dynamics.  Conservation is important as well, suggesting approaches
that implement gravitational and rotation sources conservatively, and
methods for improving angular momentum conservation.

\section{AMReX Astrophysics Suite}

Our suite of application codes is built on the
\amrex\ block-structured adaptive mesh refinement (AMR) library.  The
basic programming model has \amrex\ managing the grid data-structures
and parallel communications, and it calls the computational kernels on
a patch-by-patch basis.  The core library is written in \cpp\ with 
computational kernels written in Fortran\footnote{Currently
\maestro\ is written in pure-Fortran, but will be
 ported to the updated \cpp\ \amrex\ framework this coming year.}
---this allows us to take
advantage of the strengths of both languages. \amrex\ supports subcycling
in time, which we use in \castro.
\begin{figure}[t]
\centering
\includegraphics[width=0.58\linewidth]{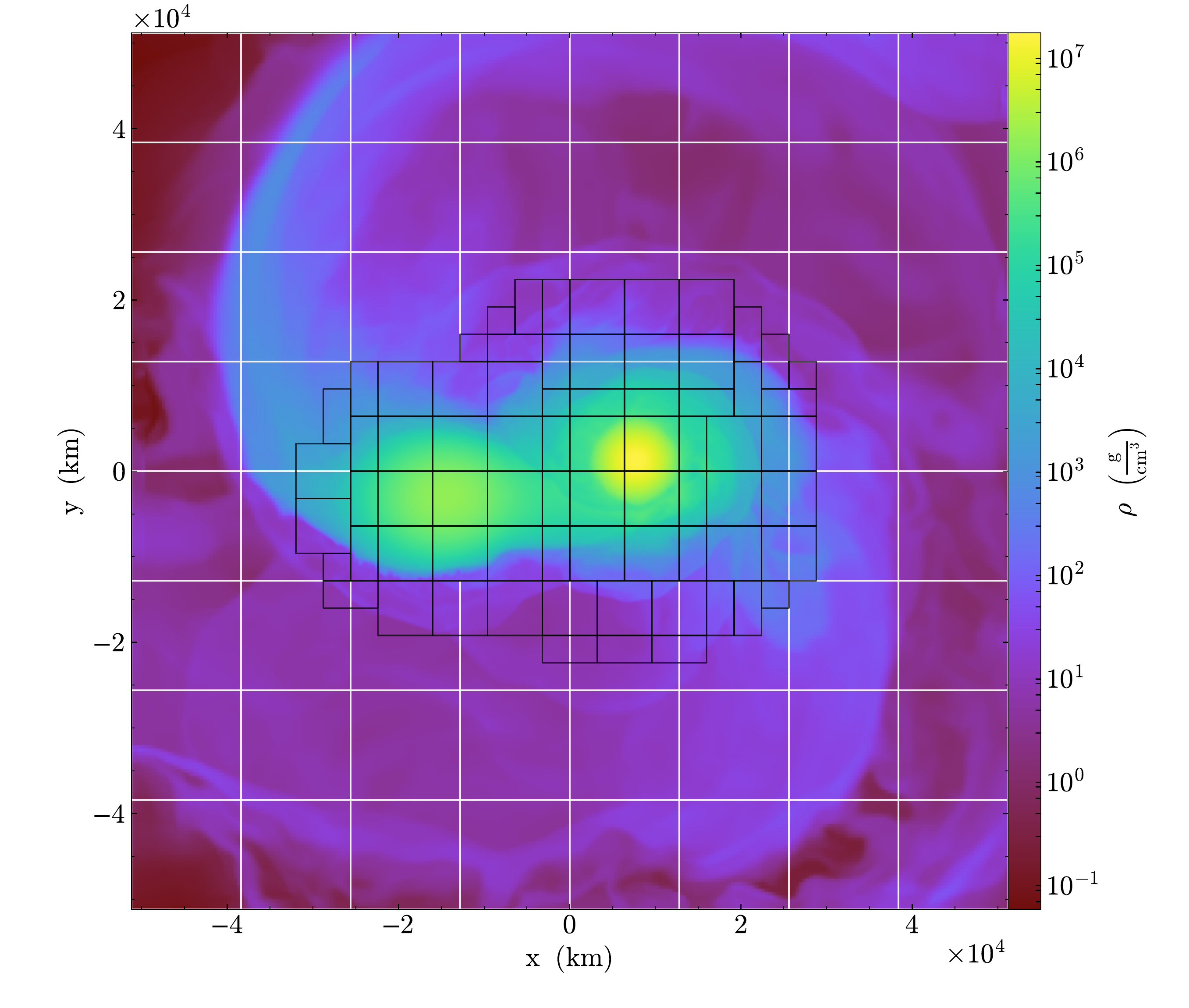}
\begin{minipage}[b]{0.38\linewidth}
\caption{\label{fig:amr_grid} A slice through a
  \castro\ three-dimensional simulation of merging white dwarfs
  showing the 2-level grid structure.  The boxes making up the base
  grid are drawn in white and the boxes making up the finer grid are drawn in black.  Note
  that the finer boxes can overlap multiple coarse boxes, and not all
  boxes are the same size.\vspace{2em}}
\end{minipage}
\end{figure}

\amrex\ uses a hybrid MPI + OpenMP approach to parallelism.
Distribution of grid patches to nodes using MPI provides a natural
coarse-grained approach to distributing the computational work, while
subdividing patches into tiles and using threads to parallelize over
tiles using OpenMP provides effective fine-scale parallelization and
amortizes thread overhead over large units of work.  Additionally,
tiling does not incur the large increase in metadata
associated with using smaller patches in a flat MPI
mode~\cite{tiling} since we can use fewer, larger patches.
This strategy is especially important for
many-core architectures like the Intel Xeon Phi.  Ongoing development is being
done to support GPU offloading, using managed memory provided by the
latest generations of GPUs.  Figure~\ref{fig:amr_grid} shows an
example of a 2-level grid.

There are many application codes built on \amrex, including those in
combustion, multiphase flow, accelerator design, and microfluidics. In
astrophysics, these include \maestro\ and \castro\ for stellar and
nuclear astrophysics applications and \nyx~\cite{nyx} for cosmological
applications.  Here we focus on the former two.

\subsection{\maestro}

\maestro~\cite{MAESTRO:Multilevel} is a low Mach number stellar
hydrodynamics code designed for efficiently modeling convection in
stars.  \maestro\ decomposes the state variables into a one-dimensional
hydrostatic base state and a three-dimensional Cartesian state that
models the deviation from hydrostatic equilibrium.
A constraint equation is derived by
requiring that the pressure everywhere is close to the background
hydrostatic pressure.  The constraint acts to enforce instantaneous
acoustic equilibration, effectively filtering soundwaves from the
system, while retaining the compressibility effects due to the
background stratification of the star and local heat release, as well
as the hydrostatic adjustment of the star.  In this fashion, it is
more general than traditional anelastic methods.  The timestep
constraint for these equations depends only on the fluid velocity, not
the sound speed, enabling much larger timesteps than compressible
codes for highly subsonic flows.

The state is advanced using a second-order accurate projection method.
Fluid quantities are advected using an unsplit Godunov method, with
reactions incorporated via operator splitting.  The provisional
velocities are then projected onto the space that satisfies the
divergence constraint.  The projections require solving a variable
coefficient elliptic equation, which is done numerically using a
multigrid algorithm.  A number of recent advances in low Mach number
modeling~\cite{kleinpauluis,vasil:2013} have been incorporated into
\maestro.

\maestro\ has been applied to convection in the Chandrasekhar-mass
model for SNe Ia \cite{ZABNW:IV,wdconvect,wdturb}, the sub-Chandra
model for SNe Ia \cite{subchandra,subchandra2},
XRBs~\cite{xrb,xrb2,xrb3}, and convection in massive stars
\cite{ms_cc}.

\subsection{\castro}

\castro~\cite{castro,castroII,castroIII} is a fully-compressible
radiation hydrodynamics code that supports arbitrary equations of
state, nuclear reaction networks, and Poisson gravity using geometric
multigrid.  The main hydrodynamics scheme in \castro\ is an unsplit
piecewise parabolic method.  The radiation solver in \castro\ uses the
flux-limited diffusion approximation for gray or multigroup radiation.
The integration algorithm on the grid hierarchy is a recursive
procedure in which coarse grids are advanced in time, fine grids are
advanced multiple steps to reach the same time as the coarse grids and
the data at different levels are then synchronized. The
synchronization for self-gravity is similar to the algorithm
introduced by \cite{miniati-colella}.  Recent developments in
\castro\ include a spectral-deferred correction method of coupling
hydrodynamics and reactions, a conservative gravity and rotation
source formulation~\cite{wdmergerI}, and a retry mechanism to redo a
step based on criteria evaluated during the
integration.

\castro\ has been applied to core-collapse
supernovae~\cite{castro-ccsne}, radiative shock breakout in
supernovae~\cite{lovegrove:2017}, population III pair-instability
supernovae~\cite{castro-pairinstability}, the Chandra model for
SNe~Ia~\cite{ma:2013}, the sub-Chandra SNe Ia model~\cite{moll:2013},
and white dwarf mergers as a model for SNe
Ia~\cite{moll:2014,wdmergerI}.  For \maestro\ simulations that evolve
from the subsonic regime to the sonic regime, we have demonstrated the
ability to restart the calculations in \castro\ to continue the
evolution into the sonic regime~\cite{scidac-petascale,malone:2014}
(in this case, for Chandra model SNe Ia).

\subsection{\starkiller\ Microphysics}

\maestro\ and \castro\ share the same microphysics, available as the
\starkiller\ Microphysics GitHub
project\footnote{\url{https://github.com/StarKiller-astro/microphysics/}}.
This includes equations of state and nuclear reaction
networks\footnote{Several of the reaction network righthand sides and
  the EOS were provided from Frank Timmes' {\em cococubed} software
  instruments page
  \url{http://cococubed.asu.edu/code_pages/codes.shtml}.  We thank him
  for making them available.}.  The reaction networks are written such
that the rates and integration strategy are decoupled, allowing us to
change the integration strategy for a given set of rates.  They are
also written to be threadsafe and with GPUs in mind (more on that
below).  The goal of \starkiller\ is to create a set of community
microphysics solvers that can be used in a variety of nuclear
astrophysics codes, not just those discussed here.

\subsection{Open source and reproducibility}

All of our simulation codes are open source and follow a fully open
development model---the development git repos are hosted on
GitHub\footnote{\url{https://github.com/AMReX-Astro/}}, available for
anyone to see and contribute to using issues and pull-requests.
Additionally, we have mailing lists for discussions and asking for
help.  Several branches are used in our workflow.  New changes are put
into the {\tt development} branch in each repo.  Nightly regression
testing is used to ensure that no new bugs were introduced.  Once a
month, {\tt development} is merged into {\tt master}.  Finally, {\em all source
  files, model files, input parameters, etc.\ for any published
  science results are also available in the code repos.}  When
feasible, the git hashes for the published results are included in
paper acknowledgments.

\section{Parallel Performance and GPUs}

A key design goal of our application codes is performance portability.
We want the same kernels to run on clusters, manycore machines (e.g.
Intel Xeon Phi), and GPU-based machines.  Our development has balanced this
need with architecture-specific optimizations to maximize code reuse.

\begin{figure}[t]
\centering
\includegraphics[width=0.48\linewidth]{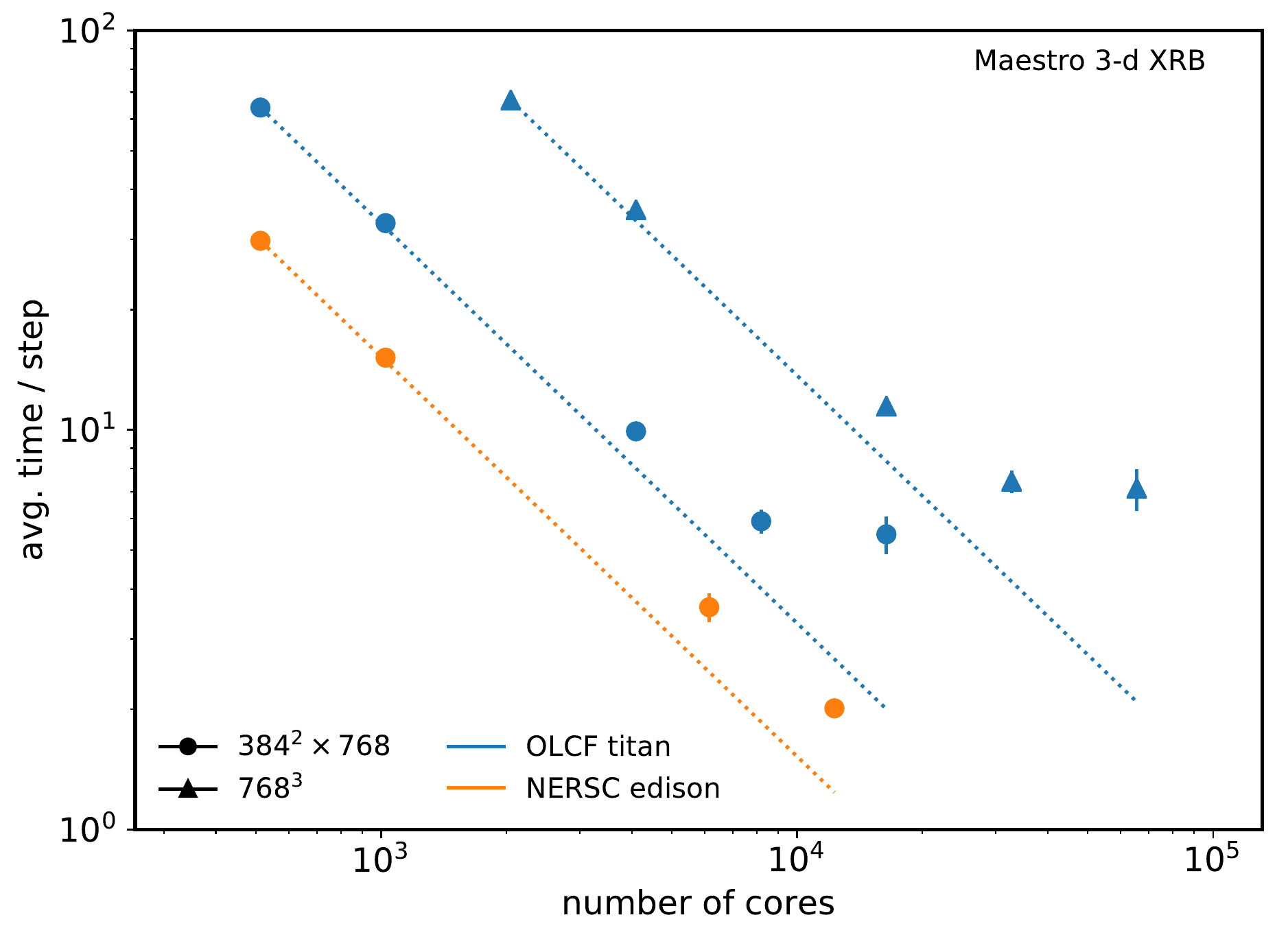}
\begin{minipage}[b]{0.48\linewidth}
\caption{\label{fig:maestro_scaling} \maestro\ strong scaling on NERSC
  Edison and OLCF Titan for a 3-d XRB problem.  Two different problem
  sizes are shown.  We see excellent strong scaling to high core counts
  for this problem.\vspace{2em}}
\end{minipage}
\end{figure}

\begin{figure}[t]
\centering
\includegraphics[width=0.48\linewidth]{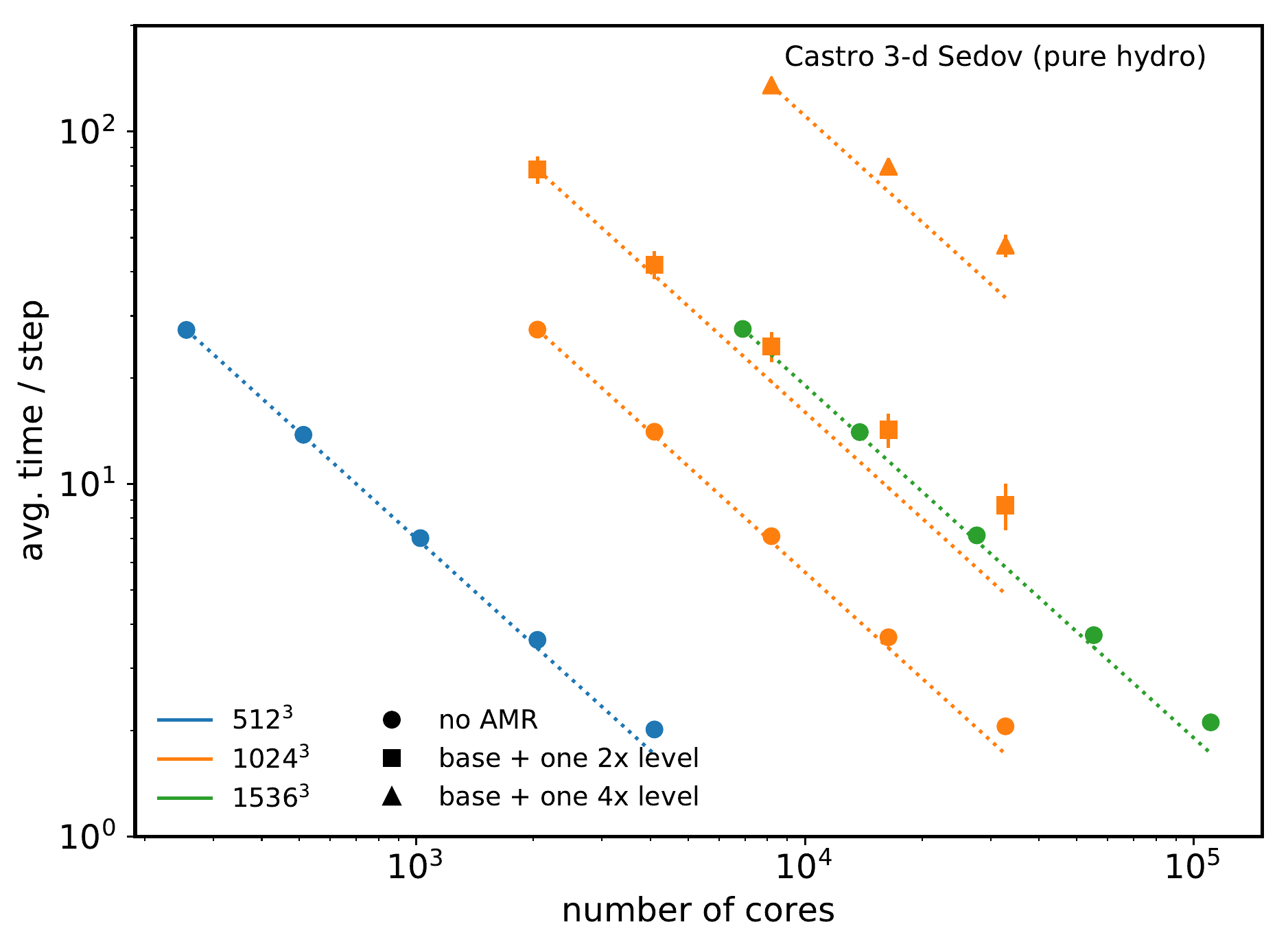}
\includegraphics[width=0.48\linewidth]{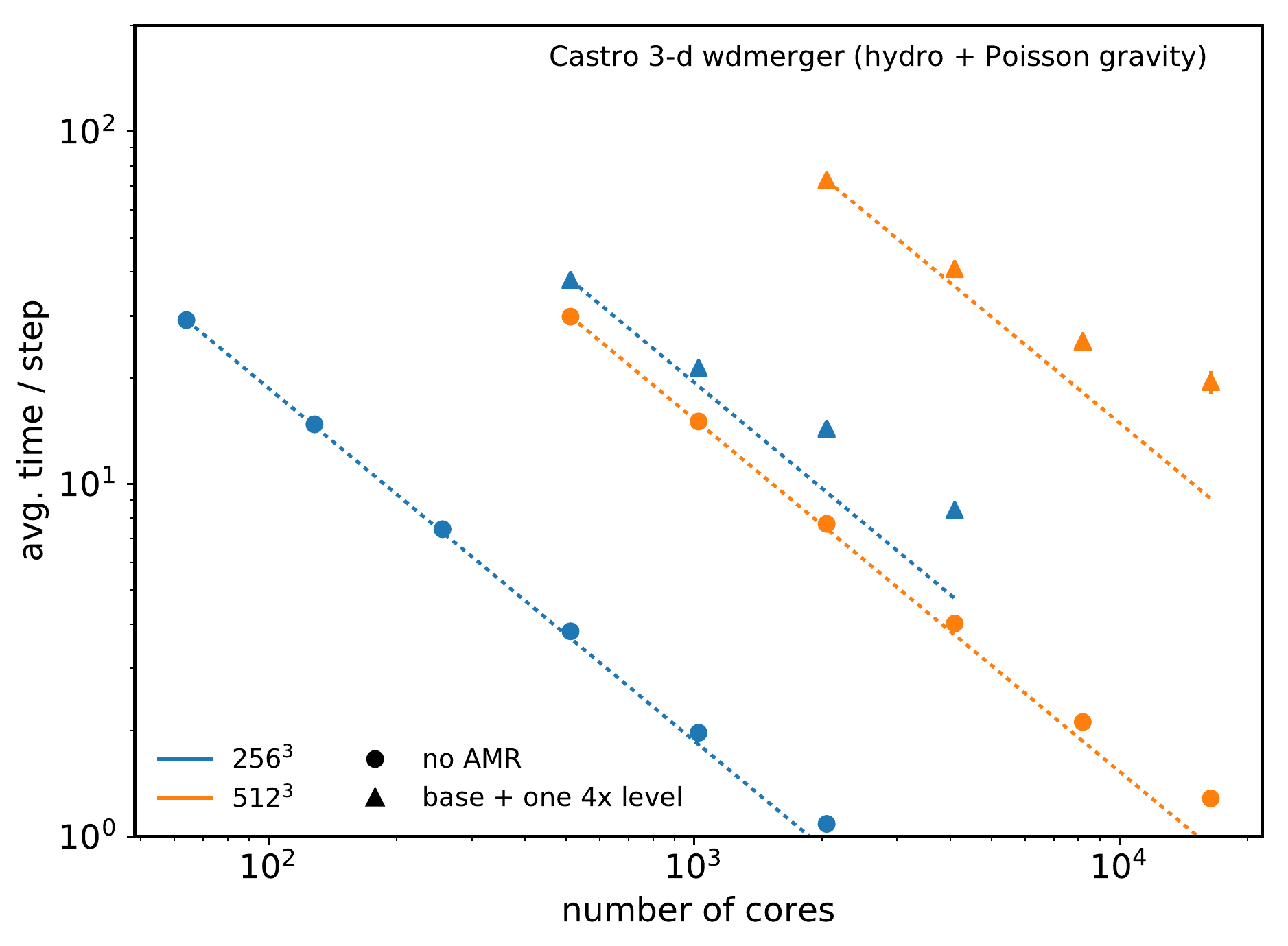}
\caption{\label{fig:castro-scaling} (left) \castro\ strong scaling on
  OLCF Titan for a pure hydro (Sedov w/ real EOS) problem.  The
  different colors represent different base resolutions.  We see
  excellent strong scaling for the single level runs.  For the
  $1024^3$ run, we also ran with one level of refinement by a factor
  of 2 (triangles) or a factor of 4 (squares), and see good strong
  scaling.  The variability (shown by the error bars) at high core
  counts shows we are becoming work-starved.  These runs used the PGI 17.7
  compilers.  (right) \castro\ strong
  scaling on OLCF Titan for a hydro + Poisson gravity (wdmerger)
  problem.  This problem uses a multipole solver to determine
  Dirichlet boundary conditions representing an isolated mass
  distribution, and then geometric multigrid to solve for the
  potential in the interior.  Two coarse grid sizes are shown, and
  demonstrate great strong scaling.  We also look at a single level of
  refinement (by a factor of 4) on top of this coarse grid.  These runs 
  used the Cray 8.5.7 compilers.}
\end{figure}

\begin{figure}[t]
\centering
\includegraphics[width=0.48\linewidth]{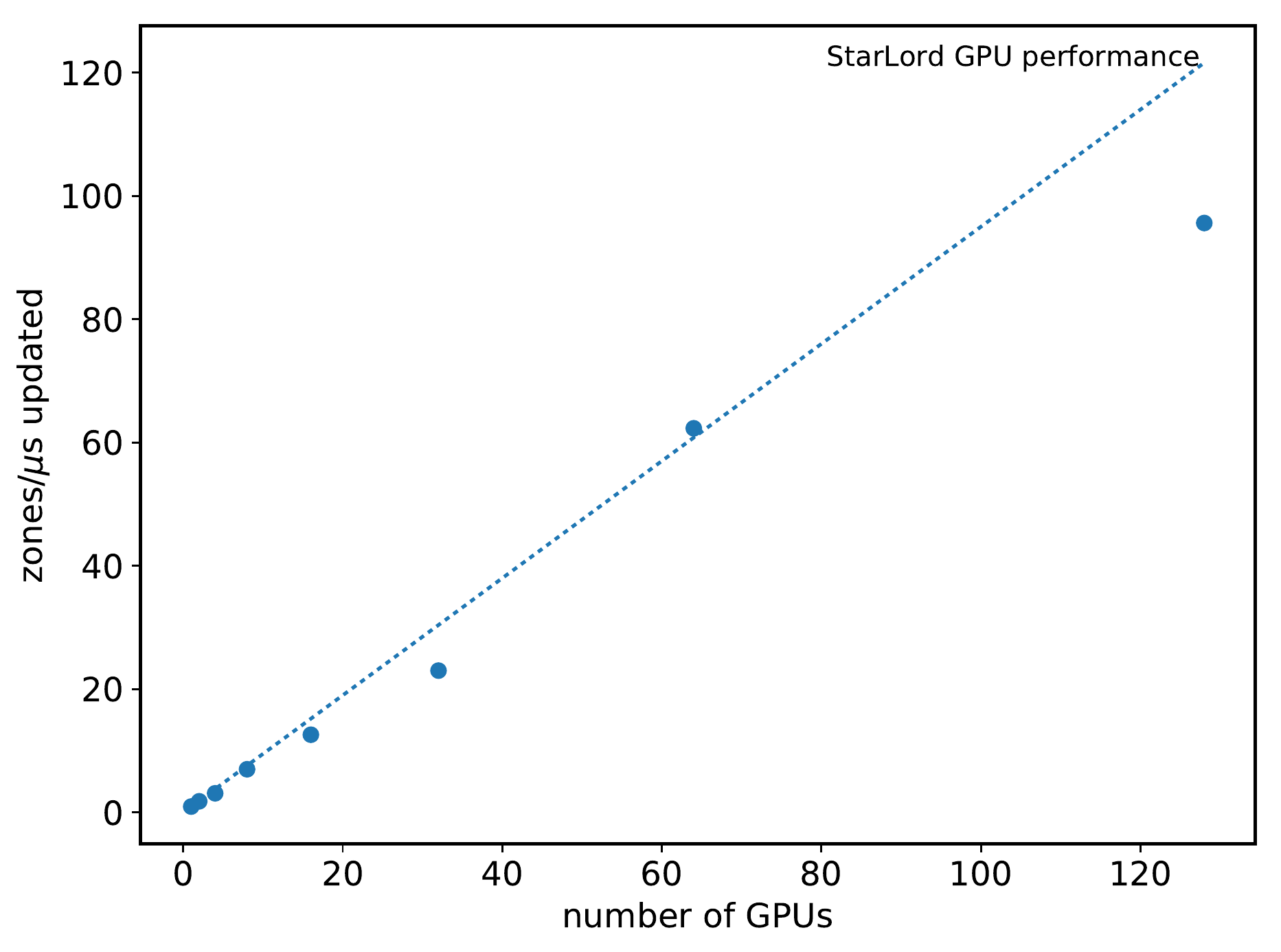}
\includegraphics[width=0.48\linewidth]{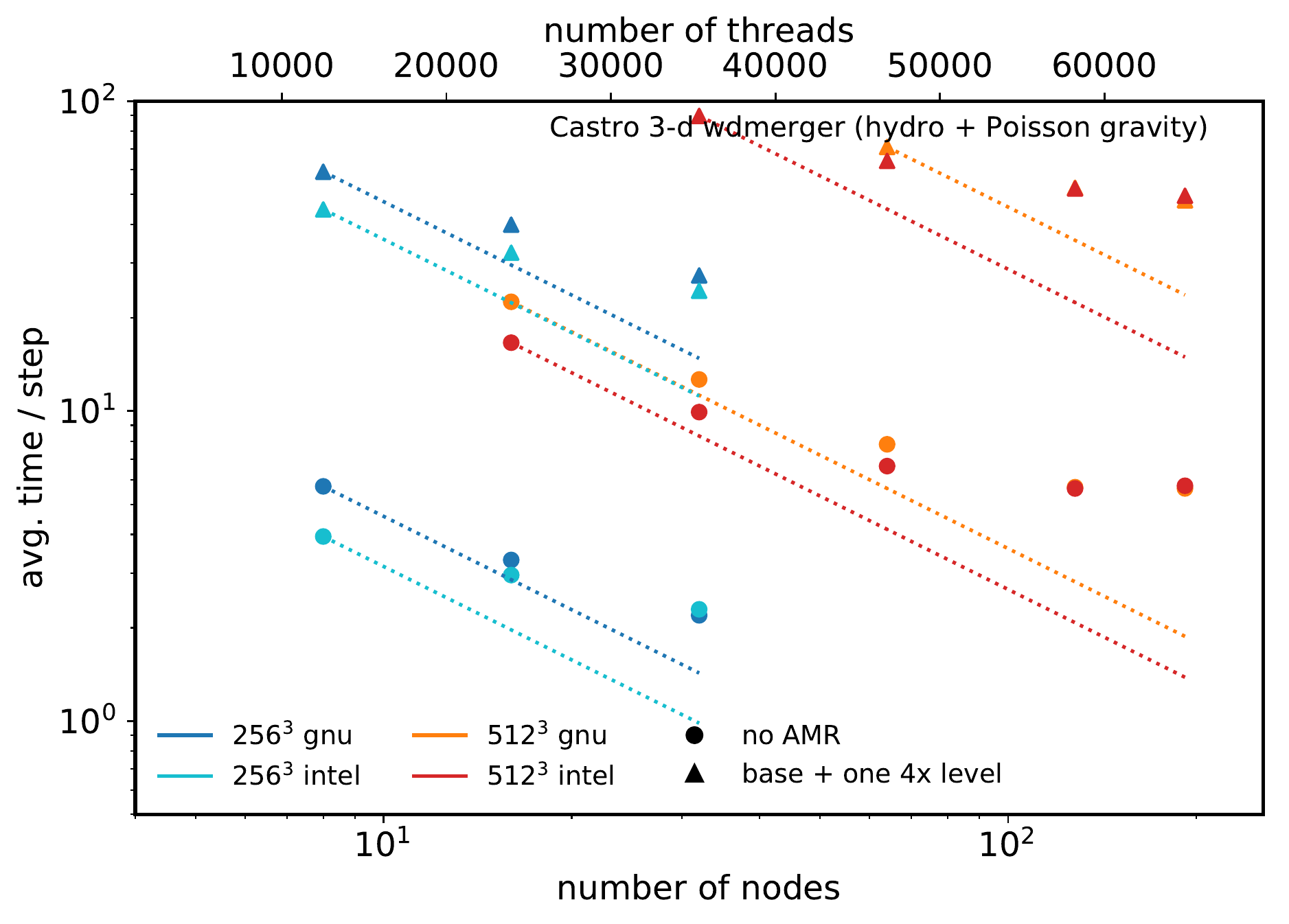}
\caption{\label{fig:knl-gpu-scaling} \castro\ scaling on new
  architectures.  The left figure shows the \castro\ GPU proxy app
  (\starlord) on OLCF Summitdev.  We see good scaling behavior to 64
  GPUs in this test of a pure hydro problem (Sedov w/ real EOS).  On the
right, we explore performance on NERSC Cori (an Intel Xeon Phi platform).}
\end{figure}

Figure~\ref{fig:maestro_scaling} shows strong scaling for \maestro\ on
the 3-d XRB problem.  This is a typical
\maestro\ application~\cite{xrb3}, where burning can be a significant
part of the overall evolution.  We ran on both OLCF Titan and NERSC
Edison.  We see that the code scales well to $\mathcal{O}(10^4)$
processors.  The upturn at the end of the scaling curves on Titan
reflect the change from 1 MPI task / 8 OpenMP threads per NUMA node to
1 MPI task / 16 OpenMP threads per compute node (2 NUMA nodes).  The
main limitation to the scaling at the moment is the multigrid solves
used to enforce the projection (in particular the nodal solver).
Also, \maestro\ currently uses a simpler fine-grained approach to parallelism
where planes in the $z$-direction are divided among OpenMP threads.
As we port \maestro\ to the \cpp\ \amrex\ code base we will take
advantage of ongoing development for increased tile size control.

Figure~\ref{fig:castro-scaling} shows \castro\ scaling behavior for a
pure hydro problem with a real EOS and a hydro + self-gravity problem
with a real EOS.  All tests were run on OLCF Titan.  The AMD
processors in Titan can be used in an 8 or 16 cores / node
configuration.  For all these runs, we ran in the 16 cores / node
mode.  We see excellent scaling behavior for the pure hydro problem to
$\mathcal{O}(10^5)$ cores.  For the self-gravity problem, we scale
very well with a uniform grid, but with a factor of 4 refinement for
the AMR level on top of this, we observe degraded performance at
higher core counts.  This is a challenging problem, as only 3\% of the
domain volume is refined.  For a given core count, there are a number
of different combinations of MPI ranks and OpenMP threads we can use.
In general, with multilevel problems, we found the best performance
with fewer MPI ranks and more OpenMP threads.

Our latest focus has been on GPU ports of our application codes.  A
small proxy app, \starlord, was created from \castro\ with just the
hydrodynamics and stellar equation of state.  It uses a simple
method-of-lines formulation of the hydrodynamics and advects 13
nuclear species in addition to the hydrodynamics.  To offload work on
GPUs, GPU support was added directly into \amrex\footnote{This support
  is currently on a feature branch in the git repo, awaiting merge
  into {\tt development}.}.  In \amrex, an iterator loops over all of
the boxes at the same level of refinement and passes a data pointer
into a Fortran kernel function where the work is done. For the
simulation state data that resides in each box, we have modified the
memory allocator so that it can use a CUDA allocator (mainly relying on
managed memory).  The domain iterator is configured to handle data
motion to and from the device, so that the compute kernels can operate
on data that is presumed to already be there, and the computation is
decoupled from the memory management. Computation on the data can then
be performed with OpenACC, CUDA Fortran, or (more recently) OpenMP
4.5. We have also built CUDA compute support into \amrex\ so that a
compute kernel can be transparently operated on using CUDA Fortran
without substantially modifying the kernels (and we anticipate using a
similar strategy to use OpenACC and/or OpenMP in the future). This
helps ensure that we can continue to maintain performance portability
in our simulation codes, by decoupling the physics algorithms from the
backend support used to implement them on various compute
architectures.

Figure~\ref{fig:knl-gpu-scaling} shows the performance of
\starlord\ on the Summitdev platform at OLCF\footnote{Summitdev
  consists of 2 IBM Power8 processors and 4 NVIDIA Pascal GPUs per
  node.}.  The highest GPU count (128) corresponds to 32 nodes on
Summitdev.  We see a nearly linear speedup with the number of GPUs,
indicating good weak scaling on this machine. A single P100 GPU achieves
a performance approximately 2.5 times that of the 20 Power8 cores. Efforts are underway to
complete the port of the GPU developments into \castro.  This figure
also shows our performance on the Intel Xeon Phi manycore processors (using
NERSC Cori).  This is for the same problem as the self-gravity test on
Titan.  The Intel Xeon Phi chips in Cori
have 68 cores than can be run with 1, 2, or 4 hardware threads each.
For all these runs, we ran with 4 threads / core (272
threads / chip).  We have not yet focused on optimizing \castro\ for the
Intel Xeon Phi architecture.

\begin{figure}[t]
\centering
\includegraphics[width=0.48\linewidth]{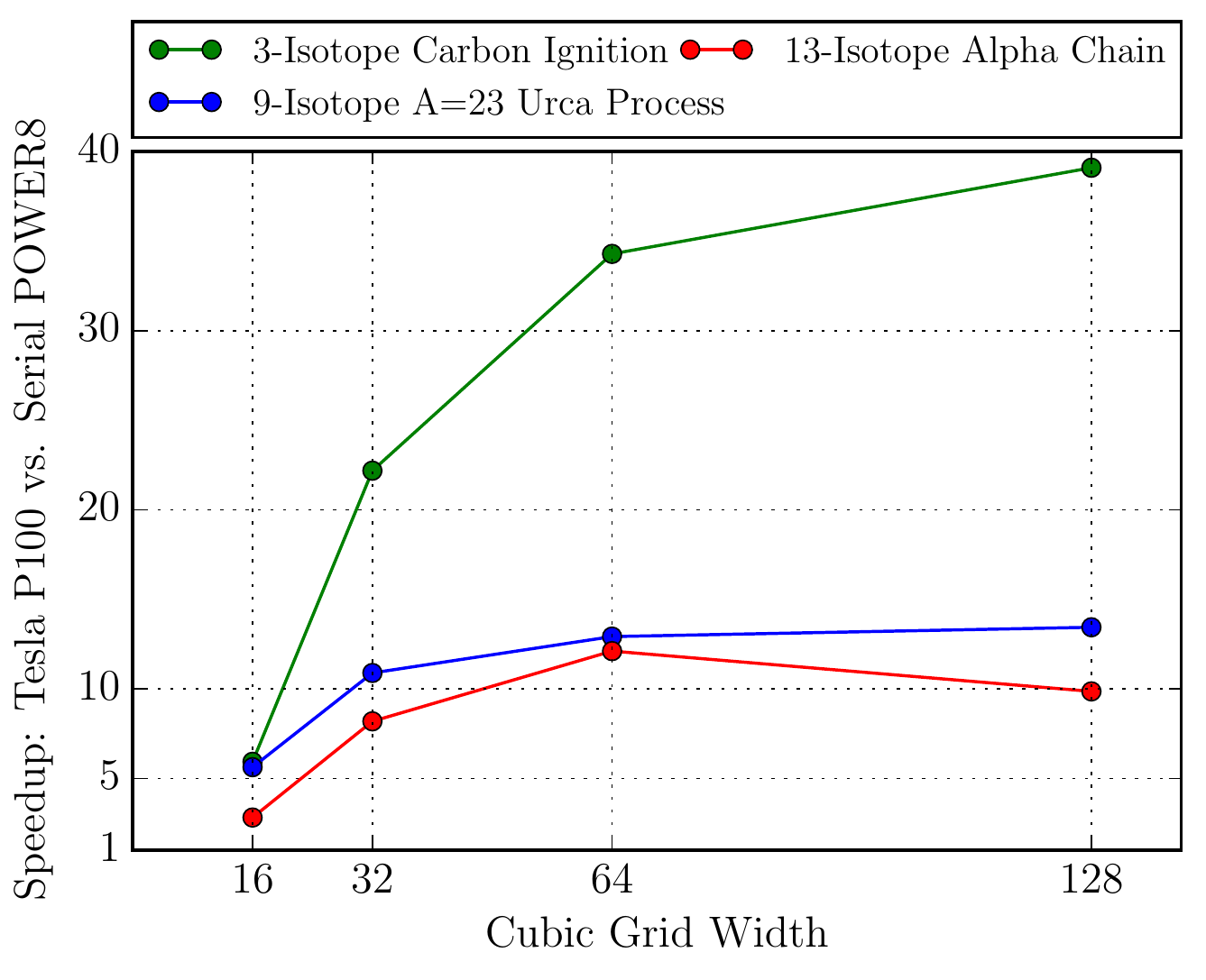}
\begin{minipage}[b]{0.48\linewidth}
\caption{\label{fig:cudaode} GPU speed up over single CPU core for three
  different reaction networks.
  reaction networks.  The 13-isotope alpha-chain is a
  potential network for the sub-Chandra simulations;
  the 9-isotope Urca network is used for our Urca simulations;
  and the 3 isotope carbon network was used for our original \maestro\
  white dwarf convection calculations.  A
  variety of grid sizes were used, $16^3$, $32^3$, $64^3$, and
  $128^3$.  In all cases, we see a good  speed up on the GPU
  vs.\ a single core.  These speed-up numbers are from the
  Summitdev machine, comparing a single Tesla P100 GPU to a single Power8 CPU core.}
\end{minipage}
\end{figure}

A parallel effort is porting our microphysics---in particular
reaction networks---to GPUs.  Our strategy is to do the entire ODE
integration on the GPU, i.e., the data for a patch of zones is passed
to the GPU, all righthand side and Jacobian evaluations and the
timestepping itself is done on the GPU, and once the burning in all
zones is completed, we access the data as needed on the CPU.  To
enable this, we ported our workhorse ODE integrator (VODE~\cite{vode})
to CUDA Fortran. This required extensive rewrites of the internals of
VODE, which was originally written using Fortran 77 syntax unsupported
by CUDA Fortran. Accelerating VODE with CUDA Fortran has proven
successful, and we now see significant performance gains with the CUDA
version of our reaction networks, even for moderate-sized networks (a
13-isotope standard network).  Figure~\ref{fig:cudaode} shows the
speed-ups on a GPU vs.\ single CPU core.  The main issue with scaling
is running out of local stack memory per thread with larger networks.
Reducing the memory footprint is a near-term goal for this work.  This
test problem also shows a lot of thread divergence due to the widely
differing thermodynamic conditions in the zones that are burning.

\section{Some science results}

\begin{figure}[t]
\centering \includegraphics[height=1.8in]{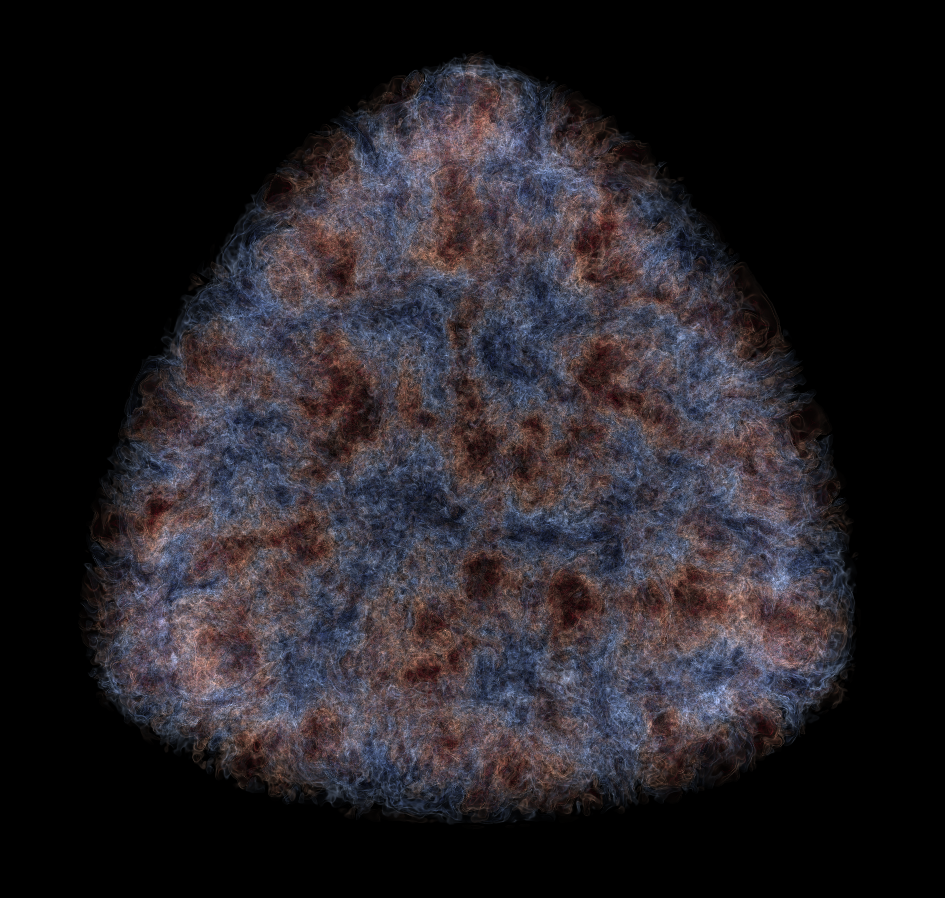}
\includegraphics[height=1.8in]{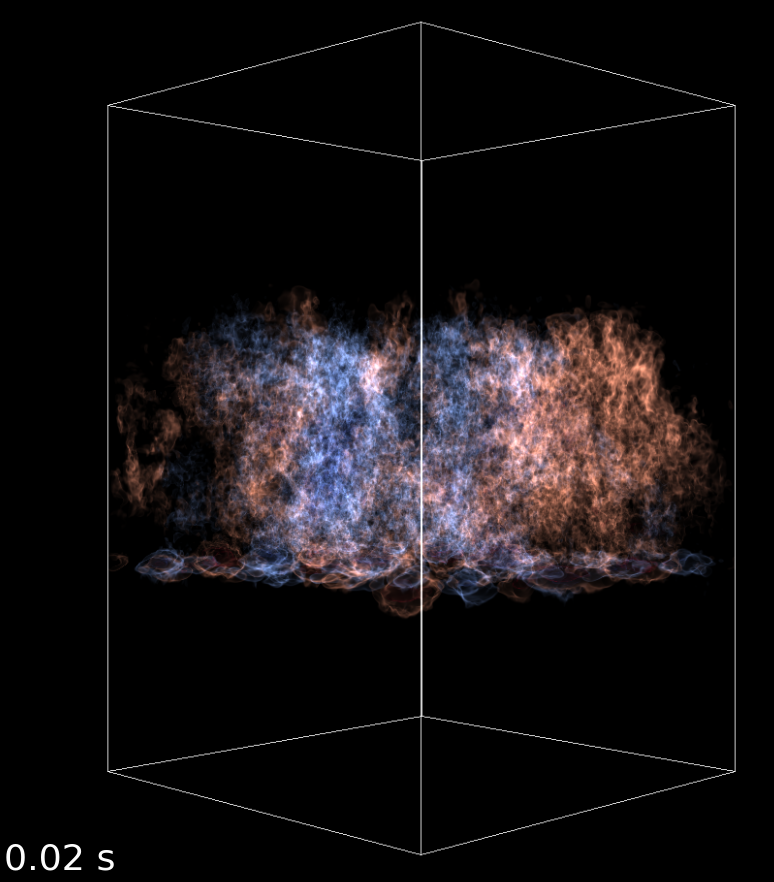}
\includegraphics[height=1.8in]{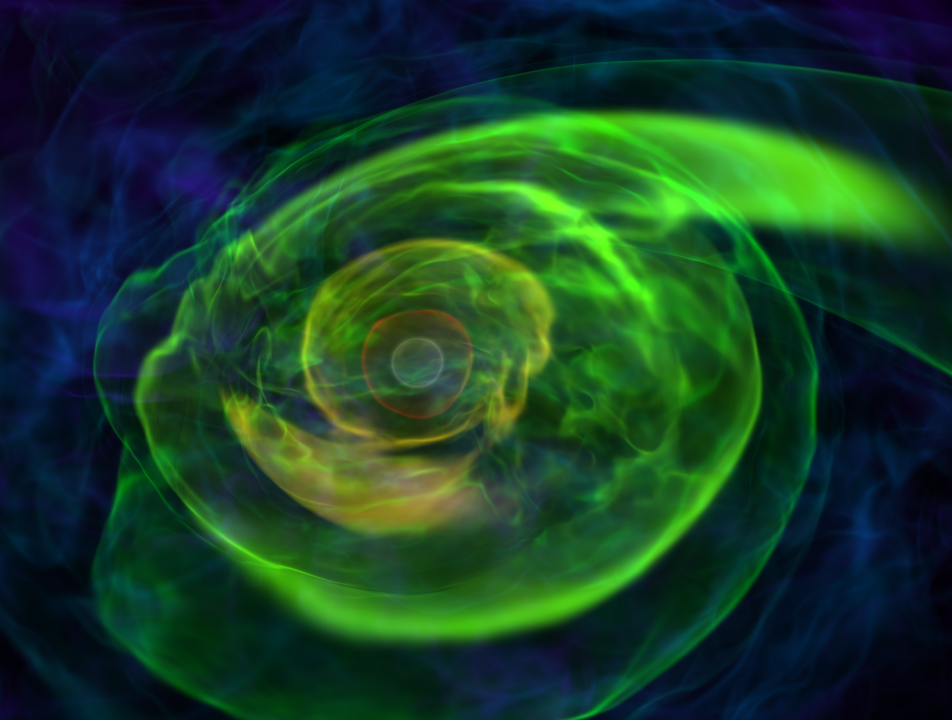}
\caption{\label{fig:current-runs} (left) Convective plumes in a
  \maestro\ sub-Ch calculation. (center) Vertical velocity showing the
  convective structure in a \maestro\ XRB calculation. (right)
  Snapshot of a \castro\ simulation of the merger of two white dwarfs,
  with 0.90 and 0.81 solar masses. The contours represent density
  levels.}
\end{figure}

Figure~\ref{fig:current-runs} shows some of our recent science
simulations.  The left panel is an image of convection in the helium
layer on a sub-Chandra white dwarf.  This is part of a study of the
early stages of the double detonation SNe Ia model.  Using \maestro,
we are able to model the convection in the He layer for many turnover
times and saw a range of outcomes depending on the mass of the white
dwarf and He layer, including a both nova-like behavior where the
entire layer runs way together and localized
burning ignited in a small region~\cite{subchandra2}.
We are performing further studies to characterize the ignition.

The middle figure shows convection in a H/He layer on a neutron star,
as a model of the early burning in an XRB.  This \maestro\ model was
the first 3D model of convection for this problem~\cite{xrb3}.  This
study showed that the convective field became fully turbulent,
achieving a Kolmogorov spectrum, and the overall dynamics was very
different than our earlier 2-d simulations.  This calculation acts as 
a bridge to our next set of studies that will look at larger scales.

The rightmost image is the coalesced remains of the merger of a
0.9~$M_\odot$ and 0.6~$M_\odot$ WD performed with \castro.  This used
the developments from \cite{wdmergerI}.  Our primary focus with this
suite of simulations is understanding the numerical sensitivity of
mergers and collisions on the burning that takes place.  This work
is ongoing.

\section{Summary and future development}

We have described our suite of astrophysics codes built on the
\amrex\ block-structured adaptive mesh refinement framework.  These
codes were developed to model problems in stellar astrophysics
spanning from low speed convection to explosive burning.  A major
theme of the codes is the open development model, with the code
development done on GitHub and all problem files needed to recreate
any science results freely available.  Future development efforts for
\maestro\ include higher-order hydrodynamics and time-integration and
rotation.  For \castro, we are investigating stronger coupling between
hydrodynamics and reactions, new solvers (including MHD), and
finishing the GPU port.

\ack The work at Stony Brook was supported by DOE/Office of Nuclear
Physics grant DE-FG02-87ER40317 and NSF award AST-1211563.  The work
at LBNL was supported by the DOE Office of Advanced Scientific
Computing Research under Contract No, DE-AC02-05CH11231. An award of
computer time was provided by the Innovative and Novel Computational
Impact on Theory and Experiment (INCITE) program.  This research used
resources of the Oak Ridge Leadership Computing Facility at the Oak
Ridge National Laboratory, which is supported by the Office of Science
of the U.S. Department of Energy under Contract
No.\ DE-AC05-00OR22725.  We appreciate the efforts of the OLCF (and in
particular Fernanda Foertter) in organizing the GPU hackathons.  This
research used resources of the National Energy Research Scientific
Computing Center, which is supported by the Office of Science of the
U.S. Department of Energy under Contract No.\ DE-AC02-05CH11231.
The authors would like to thank Stony Brook Research Computing
and Cyberinfrastructure, and the Institute for Advanced Computational
Science at Stony Brook University for access to the high-performance
LIred and SeaWulf computing systems, the latter of which was made
possible by a \$1.4M National Science Foundation grant (\#1531492).
The GPU work benefited from a grant of an Titan X Pascal board
from NVIDIA Corporation through the GPU Grant Program.
Visualizations were done using yt~\cite{yt}.  This research has
made use of NASA's Astrophysics Data System Bibliographic Services.

\section*{References}

\providecommand{\newblock}{}

\end{document}